\documentclass[traditabstract]{aa} 
\usepackage{epsfig}
\usepackage{txfonts}

\usepackage{natbib}
\bibpunct{(}{)}{;}{a}{}{,} 
\newcommand{\hi}{H\,{\sc i}}

\newcommand{\prim}{$^{\prime}$}
\newcommand{\prin}{$^{\prime\prime}$}

\newcommand{\km}{~km~s$^{-1}$}
\newcommand{\degree}{$^{\circ}$}
\newcommand{\halpha}{H${\alpha}$}

\newcommand{\msolar}{$M_{\odot}$}

\begin{document}
   \title{Galaxy evolution in Abell\,85:}

   \subtitle{I. Cluster substructure and environmental effects on the blue galaxy population}

   \author{H. Bravo--Alfaro\inst{1}
      \thanks{E-mail: hector@astro.ugto.mx}
         \and
           C.A. Caretta\inst{1}
         \and
           C. Lobo\inst{2,3} 
         \and
           F. Durret\inst{4}
         \and
           T. Scott\inst{5}
          }

\institute{Departamento de Astronom\'{\i}a, Universidad de Guanajuato.
Apdo. Postal 144. Guanajuato 36000, M\'exico
      \and
Centro de Astrof\'{\i}sica da Universidade do Porto, Rua das Estrelas
4150-762 Porto, Portugal.
       \and
Depto. de Matem\'atica Aplicada, Faculdade de
Ci\^encias, Univ. do Porto, R. do Campo Alegre 687, 4169-007 Porto, Portugal
      \and
Institut d'Astrophysique de Paris, CNRS, UMR 7095, Universit\'e
Pierre et Marie Curie, 98bis Bd Arago, 75014 Paris, France
      \and   
Center for Astrophysics Research, University of Hertfordshire, 
College Lane, Hatfield, AL10 9AB, UK
          }

   \date{Received 31 July 2008; accepted }

 
   \abstract {In this series of papers we explore the evolution of
   late-type galaxies in the rich cluster Abell\,85.  In this first
   paper we revisit the complex dynamical state of A\,85 by using
   independent methods.  First, we analyze the galaxy redshift
   distribution towards A\,85 in the whole range 0--40,000\,\km, and
   determine the mean redshifts of the background clusters A\,87 and
   A\,89, very close in projection to A\,85.  Then we search for
   substructures in A\,85 by considering the 2-D galaxy distribution
   of its members (13,000--20,000\,\km) and by applying the
   kinematical 3-D $\Delta$-test to both projected positions and
   radial velocities.  This clearly reveals several substructures: one
   close to the cluster core and three more projected towards the
   southeast, along the region where an X-ray filament has been
   extensively studied.  We also analyse the distribution of the
   brightest blue galaxies across a major fraction of the cluster
   volume, considering if they are gas-rich or poor.  We report a very
   asymmetric distribution of the blue member galaxies, with most of
   them to the east and southeast, namely in the region joining the
   core of A\,85 to its farthest substructure in this direction -
   dubbed the \emph{SE} clump.  By matching our sample of bright blue
   member galaxies with \hi\ detections reported in the literature, we
   identify gas-rich and gas-poor ones. As expected, the \hi-rich blue
   galaxies follow the same trend as the parent sample, with most of
   them projected on the eastern side of the cluster as
   well. Interestingly no blue objects have been detected in \hi\ up
   to a projected radius of 2\,Mpc in this zone. We finally estimate
   the ram pressure stripping exerted by the intra-cluster medium as a
   function of the projected distance from A\,85, in order to quantify
   how important this mechanism might be in sweeping the gas out of
   the infalling spirals.}

   \keywords{Galaxies: evolution - Galaxies: clusters: individual: Abell\,85
            }

   \titlerunning{Substructures in Abell\,85}
   \authorrunning{Bravo-Alfaro et al.}
   \maketitle
%

\section{Introduction}\label{sec_intro}

Important evolution is known to occur between $z \sim 1$ and the
present time. Galaxy clusters keep evolving throughout these epochs by
accreting galaxies and groups of galaxies, mainly along filamentary
structures \citep[e.g.][and references therein]{adami05}, which gives
rise to merging episodes in a hierarchical fashion. Many clusters
display virialized cores surrounded by infall regions of galaxies
and/or groups, which are gravitationally bound but not in equilibrium
with the main cluster body. The presence (or absence) of such
substructures is relevant since it provides tight constraints on
models of structure formation and evolution at large scales. Important
changes are also evident for individual galaxies: the cosmic star
formation rate may suffer a change of an order of magnitude since $z =
1$ \citep[e.g.][]{lilly96,madau96}. Further major evidence of galaxy
evolution is the decreasing number of spirals as one goes from the
field to the central regions of clusters -- and the corresponding
rising population of lenticulars -- in the nearby Universe \citep[the
morphology-density relation,][ and references
therein]{oemler74,dressler80,dressler04}.

Nowadays an interesting debate has arisen about this last topic: while
some authors discard major environmental effects on structural
parameters such as size and mass concentration \citep{li06}, extensive
multifrequency surveys over large volumes of galaxy clusters have, on
the other hand, provided new insights into the environmental effects
on both star formation activity and morphology of individual galaxies
\citep[see for example][]{cayatte90,vangork04,vollmer04,chung07}. 
Furthermore, the presence of groups of E+A galaxies in very specific 
regions of clusters like Coma strongly suggests that they have 
suffered recent -- and structural -- environmental processing.

In this work we revisit the dynamical state of Abell~85 (hereafter
A\,85) and the strength of ram pressure stripping to produce the gas
deficiency observed to large distances from the cluster center.  In a
forthcoming paper (Paper II) we will study the possible gravitational
effects exerted on individual galaxies infalling from the SE, by
considering their gas content, activity features and stellar
morphology seen in NIR. The goal of this series of papers is to
explore the transformation of late-type galaxies as a function of the
cluster environment under different scenarios such as effects produced
by the intra cluster medium (ICM) and/or gravitational preprocessing.

The richness class 1 cluster A\,85, located at redshift $z = 0.055$,
is a very interesting system that provides the perfect laboratory for
studying the effects of environment on the galaxy population and on
the properties of the galaxies themselves (gas content, SFR, AGN
activity, etc). This cluster has been extensively studied in the radio
continuum, optical and X-rays
\citep[e.g.][]{durret98b,giovannini99,kempner02,durret05}.  Many works
have revealed a dynamically young system undergoing several mergings.
For example, \citet{kempner02} proposed the existence of two
subclusters around the inner part of A\,85, and studied one of them
based on Chandra observations; they report a subcluster much less
massive than A\,85, merging with the latter from the south.  Two other
Abell clusters, A\,87 and A\,89, are very close in projection to A\,85
\citep{abell89}.
\citet{durret98b} concluded that A\,89 is rather constituted by two
\emph{sheets} of galaxies on the background, while A\,87 would not be a
separate cluster but rather a series of individual groups moving into
the main body of A\,85 from the SE. The global cluster orientation, as
traced by X-rays and by the optical galaxy distribution, indicates an
elongation towards the A\,87 region, at a position angle of
160\degree, coincident with the large scale structure in which the
cluster is embedded. This elongation also coincides with the
orientation of the southern X-ray filament seen in different surveys.

We follow a systematic strategy to detect substructures around A\,85
based on a thorough analysis applied to a large set of optical
photometric and spectroscopic data (described in
Sect.\,\ref{sec_data}). Our first approach is based on revisiting the
galaxy redshift distribution in the direction of A\,85, up to
40,000\,\km ($z \sim 0.133$), and identifying the different systems
along the line of sight (Sect.\,\ref{sec_vel}).  We then probe, in
Sect.\,\ref{sec_2D}, the 2-D spatial distribution of the member
galaxies of A\,85 in the same field of view.  For a better assessment
of three-dimensional sub-structures, we perform a 3-D kinematical
analysis of the cluster making use of the $\Delta$-test
\citep{dressler88}, which is one of the most sensitive algorithms to
trace substructures.  This is described in Sect.\,\ref{sec_3D}.  As a
complementary method for probing substructures, we carry out an
analysis of the distribution of the brightest blue galaxies in A\,85
(in Sect.\,\ref{sec_bluegals}),
considering if they are \hi-rich or poor by matching positions 
with \hi\ detections reported previously in the literature. 
We discuss in Sect.\,\ref{sec_discuss} the results yielded 
by the different methods and trace a scenario of galaxy evolution 
in A\,85 where ram pressure seems to have played an important role 
in the evolution of the bright blue galaxies. This is supported by 
our estimates of the strength of ram pressure stripping as a 
function of the projected distance from the cluster center. 
We summarize our results in Sect.\,\ref{sec_sum}.

Throughout this paper we assume $\Omega_M = 0.3$, $\Omega_\Lambda =
0.7$, and $H_0 = 75$ km\,s$^{-1}$\,Mpc$^{-1}$.  In this cosmology, the
scale is 10\prim\ = 0.6\,Mpc at the cluster mean redshift.


\section{Observational data}\label{sec_data}

We obtained position and optical magnitude data for galaxies in the
A\,85/87/89 complex from the SuperCOSMOS Science Archive, available
from the Royal Observatory of
Edinburgh\footnote{http://surveys.roe.ac.uk/ssa/index.html}.  The area
covers a region of 50 arcmin radius around the center of A\,85, taken
as the position of its cD galaxy, MCG-02-02-086, which also coincides
with the peak of the X-ray emission of the cluster.  Magnitudes in
$b_{\rm J}$ and $r_{\rm F}$ bands came from the digitization of UKST
Survey plates (IIIa-J + GG395 and IIIa-F + OG590, respectively) by the
SuperCOSMOS machine, using 10$\mu$m size pixels (0.67'')
\citep{hambly01a}. These magnitudes were placed on an absolute scale
using SDSS-EDR \citep{stoughton02}, EIS \citep{arnouts01} and
ESO-Sculptor Survey \citep{arnouts97} data, and are good to about 0.1
mag (rms).  Note that a new calibration has been applied that
supersedes the original SuperCOSMOS Sky Survey.  The detection limit
of SuperCOSMOS data is estimated to be about $b_{\rm J} = 20.5$ and
$r_{\rm F} = 19.5$, at a completeness level above 95\%, while the
astrometry is accurate to better than 0.3\prin\ in both right
ascension and declination \citep{hambly01b}.

This photometric catalogue was then matched with radial velocity data
from the compilation of \citet{andernach05}\footnote{Most of the data
for A\,85 in this compilation comes from
\citet{bee91,malu92,durret98a,christlein03,smith04} and SDSS
\citep{stoughton02}}, the largest compilation of velocities for Abell
cluster member galaxies to date. For the A\,85/87/89 complex (the area
inside a radius of 50 arcmin around the center of A\,85) this
compilation contains 1693 entries, including multiple observations for
most of the members.  After discarding measurements with large
deviations we averaged the radial velocities for each galaxy.  This
resulted in redshifts for 574 galaxies in the area of the A\,85/87/89
complex in the range 0--40,000\,\km.  In the range
13,000--20,000\,\km, corresponding to A\,85
\citep[e.g.][]{durret98a}, the compilation yields 367 redshifts 
for member galaxies.  We found a match of 98.9\% between these
galaxies and SuperCOSMOS data.

In order to study the global \hi\ content in A\,85 we consider the
positions of twelve \hi-detected galaxies in this cluster recently
reported by \citet{hba08}. This was obtained from an \hi\ survey of a
region of about 1.5$\times$1.5 degrees centered on A\,85, observed
with the VLA in its C-configuration, pointing the array at multiple
positions and central frequencies. In this fashion the whole complex
A\,85/87/89 has been homogeneously surveyed with an average noise of
rms$\sim$0.25\,mJy\,beam$^{-1}$, within the velocity range
14,580--18,440\,\km.  These observations have, on average, an \hi\
mass detection threshold of 7$\times$10$^8$\,\msolar, corresponding to
a 6$\sigma$ detection level.  More details on the observations, on the
complex data reduction process, and the gaseous properties of
individual galaxies will soon appear (van Gorkom et al., in prep.)

%
\section{Results on the galaxy distribution}\label{sec_res}

\subsection{The velocity distribution towards A\,85}\label{sec_vel}

\begin{figure}
\resizebox{\hsize}{!}{\includegraphics{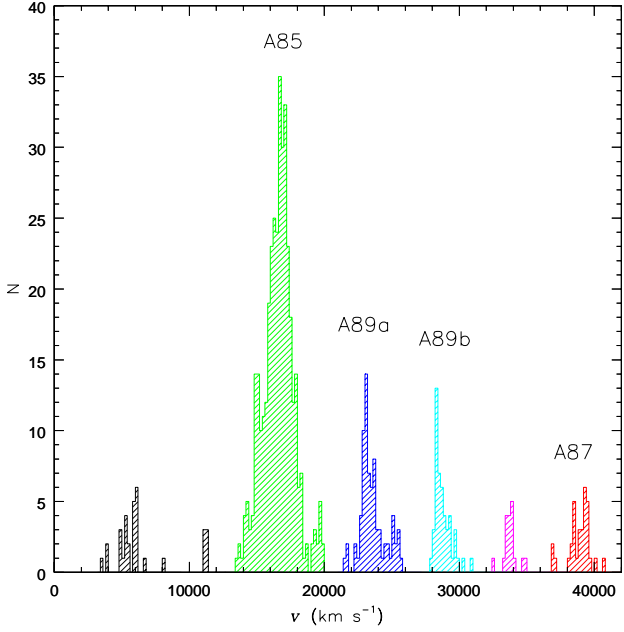}}
\caption{Histogram showing the distribution of 574 velocities in 
the region of the A\,85/87/89 complex. The most conspicuous concentrations 
are marked according to their cluster and/or group association.}
\label{histogr_0_40}
\end{figure}

\begin{figure}
\resizebox{\hsize}{!}{\includegraphics{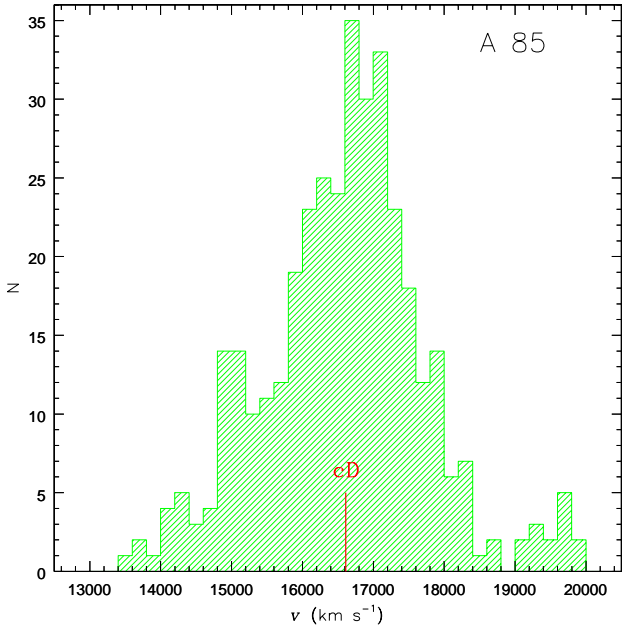}}
\caption{Histogram showing the velocity distribution of galaxy-members of
A\,85, within the range 13,000--20,000\,\km.}
\label{histogr_13_20}
\end{figure}

We revisit the velocity distribution in the direction of A\,85 by
analyzing the redshift catalogue compiled in Sect.\,\ref{sec_data}.
The distribution of these radial velocities is shown in
Fig.\,\ref{histogr_0_40}. As expected there are many concentrations of
galaxies along this line of sight. The most remarkable one is the
A\,85 spike, within the velocity range 13,000-- 20,000\,\km. This peak
corresponds to 64\% of the galaxies in the cone.  We find a biweight
mean radial velocity of 16,583 (+60, -52)\,\km\ and a velocity
dispersion of 1,122 (+54, -54)\,\km\ for the 367 member galaxies in
this system, using a robust statistical ({\sc rostat}) estimation
\citep{bee90}.  Their velocity distribution, enlarged in
Fig.\,\ref{histogr_13_20}, is not far from a Gaussian.  It shows a
small positive skewness of 0.11, probably because of the small
sub-peak between 19,000 and 20,000\,\km.  Most of the galaxies in this
velocity sub-peak have positions coincident with a subcluster located
in the western region of A85, that may be infalling into the main
cluster.  The kurtosis of the distribution is also positive (0.49),
revealing a slightly peaked shape.

The next two prominent peaks along the line of sight to A\,85, at
23,500 and 28,500\,\km\ respectively, whose centers are projected in
the direction of A\,89 (especially the former one), refer to the near
and remote sheets identified by \citet{durret98b}.  Our sample
consists of 77 and 48 galaxies for these two structures,
respectively. The last peak ($\sim$39,000\,\km) is centered on the
A\,87 region. This is most probably the system of galaxies associated
with the A\,87 cluster identified by \citet{abell89}.  A\,87 is a
distance class 5 cluster, which, following Abell's estimation,
corresponds to about 39,000\,\km\ on average ($z \sim 0.131$). With 32
member galaxies, the {\sc rostat} calculation gives a biweight mean of
39,014 (+113, -143)\,\km\ and a velocity dispersion of 737 (+259,
-110)\,\km\ for this system of galaxies.

In general, our values of central velocity and velocity dispersion are
in good agreement with recent works, for instance that of
\citet{christlein03}. In spite of a slight difference in the
velocity range we consider (13,000--20,000\,\km\ against
13,423--19,737\,\km\ in their work), our systemic velocity
(16,583\,\km) is consistent with theirs (16,607\,\km). Instead, a
small difference is seen between our reported velocity dispersion
(1,122\,\km) and Christlein \& Zabludoff's value (993\,\km), likely
due to our larger velocity range and sample of galaxies (367 in this
work against 280 in theirs).

%

\subsection{The 2-D galaxy distribution in A\,85}\label{sec_2D}

\begin{figure}
\resizebox{\hsize}{!}{\includegraphics{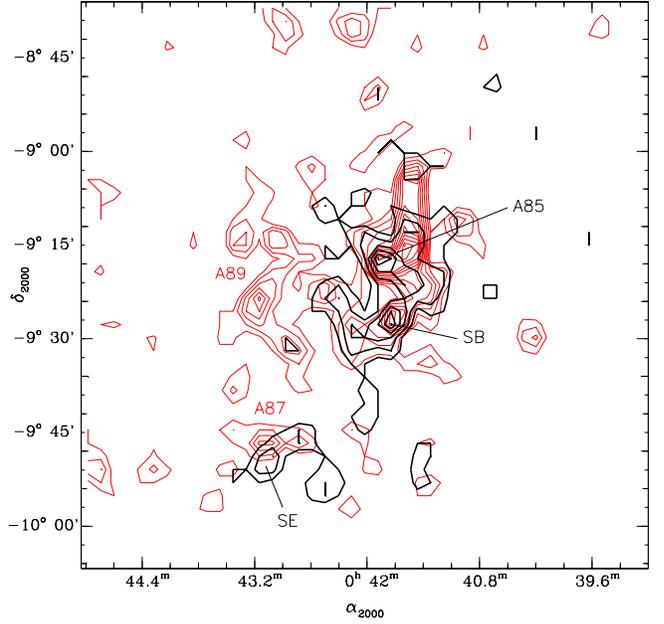}}
\caption{Iso-density contours drawing the distribution of galaxies in the
direction of A\,85; red thin contours in the background correspond to 
all galaxies with magnitudes $b_{\rm J} < 20.5$. Black thick contours indicate 
the distribution of the 367 members of A\,85 (13,000-20,000\,\km). 
Main clusters and substructures of A\,85 are indicated.}
\label{isodensity}
\end{figure}

To better understand the structure of A\,85 we constructed an
isodensity contour map for the cluster members, i.e., those 367
galaxies with radial velocities in the range 13,000--20,000\,\km.
This is shown in Fig.\,\ref{isodensity} with black contours.  The
galaxies were counted within square cells of 4.1\prim \ side, each
cell displaced only half of its size from the previous one in order to
smooth the contours. This provides a resolution of about 0.125\,Mpc at
the distance of A\,85.  The plotted contours correspond to 3, 5, 7, 9
and 11 galaxies per cell (the lowest level corresponds to 0.2
galaxies/arcmin$^2$, while the peak reaches 0.8 galaxies/arcmin$^2$).
These contours are superimposed on the isodensity map of all galaxies
with magnitudes $b_{\rm J} < 20.5$ in the A\,85/87/89 complex area
(red contours).  The A\,89 concentration appears only in red contours,
whereas the A\,87 region presents a superposition of two
structures. These structures are the A\,87 cluster (at 39,000\,\km,
seen in red contours) and, slightly displaced to the south, an
infalling subcluster of A\,85 (seen in black contours and labeled
\emph{SE} in Fig.\,\ref{isodensity}), the latter coincident with one
of the clumps found by \citet{durret98b} in the filament region.

The black isodensity contours constitute the first galaxy density map
of A\,85 built exclusively with galaxy members.  It reveals three
significant peaks: the first represents the core of the main cluster
(indicated with \emph{A85} in the figure), nearly coincident with the
cD galaxy; a secondary peak is coincident with the Southern Blob
\citep[\emph{SB} as in][]{kempner02}; and a third, and less
conspicuous one, appears some 35\prim\ (about 2.1\,Mpc) to the
south-east of the cluster center (indicated as \emph{SE} in the
figure), which corresponds to the subcluster superposed on A\,87
already mentioned above.  We analyze in more detail this \emph{SE}
subcluster in the next section.

%

\subsection{The 3-D $\Delta$-test}\label{sec_3D}

\begin{figure}
\resizebox{\hsize}{!}{\includegraphics{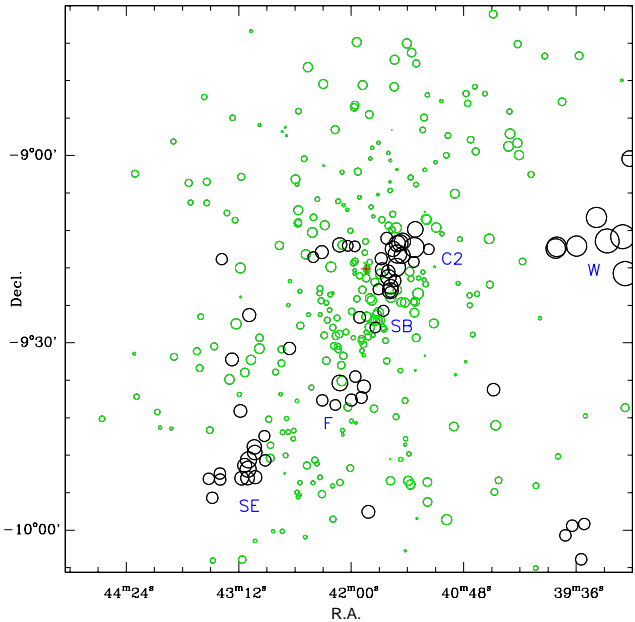}}
\caption{Substructures detected around A\,85 following the kinematical analysis
of the $\delta$ parameter. Larger black circles indicate higher probabilities
of the presence of substructures. The most conspicuous substructures detected
are indicated. The red cross marks the cluster center, coincident with the
cD.} 
\label{delta}
\end{figure}

As pointed above the galaxy density map of A\,85 members with
velocities between 13,000 and 20,000\,\km\ already shows a non-uniform
distribution.  Now we move a step forward in quantifying the presence
of substructures by applying a more powerful statistical test that
considers both positions and velocities. One of the most sensitive
algorithms proposed in the literature \citep[e.g.][]{pinkney96} is the
$\Delta$-test proposed by \citet{dressler88}, which searches for
deviations of the local velocity mean and dispersion from the global
values.  These statistics produce a deviation measure ($\delta$
parameter) for each galaxy, which can be used to find subclumps of
``deviant'' galaxies:

\begin{equation}\label{eq_delta}
\delta^2 = \frac{(N_{neighb} + 1)}{\sigma^2} [(v_{local} - v)^2 + (\sigma_{local} - 
\sigma)^2]
\end{equation} 
where $N_{neighb}$ is the number of neighbors used to estimate the
local mean velocity ($v_{local}$) and velocity dispersion
($\sigma_{local}$), while $v$ and $\sigma$ are the global mean
velocity and velocity dispersion, respectively.

A correlated spatial and kinematic variation, i.e., the presence of a
region (R.A., Dec.) containing several galaxies that deviate from the
global kinematic values, indicates a high probability of substructure
in this zone. In the following discussion we indicate these clumps of
galaxies as showing high \emph{kinematical deviations} from the global
behavior. The cumulative value of $\delta$ for the entire cluster
($\Delta$) can be compared to Monte Carlo simulations to estimate the
confidence level of the general deviation of the cluster.

We applied the $\Delta$ statistics to the 367 members of A\,85, using
$N_{neighb} = 10$, and the corresponding results are shown in
Fig.\,\ref{delta} (we obtained similar solutions with values of
$N_{neighb} =$ 15, 19 $\simeq \sqrt N$, and 37 $\simeq N/10$).  In
Fig.\,\ref{delta} each galaxy is marked by a circle with a size
proportional to $\delta$ (instead of $e^{\delta}$, as usual, since the
signal in A\,85 is very high).  We take the arbitrary value
$\delta$=2.0 as a convenient threshold to discriminate objects with
kinematics that deviate significantly from the cluster one. Therefore,
galaxies with $\delta > 2.0$ are indicated with black circles in
Fig.\,\ref{delta}, while green circles represent galaxies with $\delta
< 2.0$. The cumulative $\Delta$ parameter for the cluster is larger
than the one obtained in 99.3\% of 1,000 Monte Carlo random
simulations.

We note five prominent regions displaying enhanced variation in
kinematics with respect to the mean; these are spots of larger black
circles in Fig.\,\ref{delta}: the first one is close to the center, a
few arcmin to the W, and we identify it as \emph{C2}. Another
substructure is found towards the south, coincident with the Southern
Blob (\emph{SB}), and two more appear in the southeast region of
A\,85: one along the X-ray filament detected in previous works, namely
with ROSAT and XMM-Newton data
\citep[\emph{F}, see][and references therein]{durret03}, and the other 
along an extension of the latter (indicated with \emph{SE}). We
detect a last substructure to the west (\emph{W}) of the cluster.

%

\subsection{The distribution of the brightest blue galaxies
in A\,85}\label{sec_bluegals}

\begin{figure}
\resizebox{\hsize}{!}{\includegraphics[angle=0,scale=1.0]{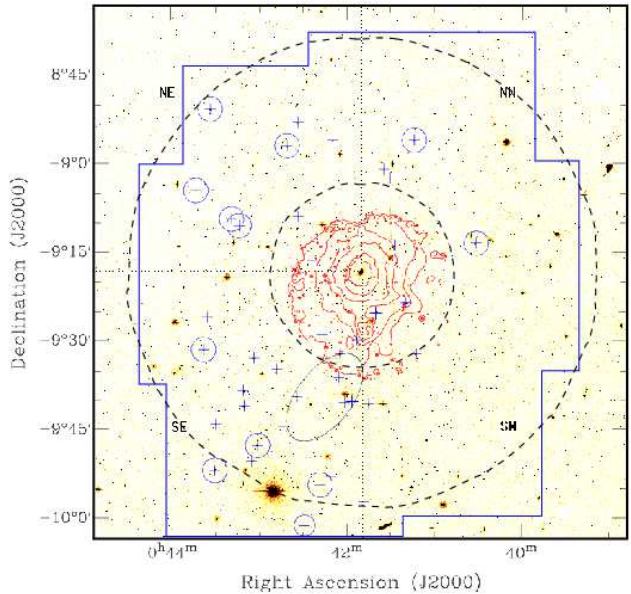}}
\caption{Distribution of the brightest blue galaxies through A85,
indicating the position of the \hi\ rich (encircled crosses) and the
\hi\ poor galaxies (crosses). The center of the X-ray emission (red 
contours) as seen by ROSAT coincides with the cD galaxy. The 
ellipse roughly draws the position of the X-ray filament mentioned in the text. 
Large dashed circles correspond to the 1.0\,Mpc and 2.4\,Mpc radius. 
The blue external polygon indicates the VLA-\hi\ surveyed area.}
\label{blueglobal}
\end{figure}

Another effective method to trace substructures in clusters consists
of analysing the distribution (both in projected position and
velocity) of their brightest blue members in a first step, and then
verifying their gas content.  This method has successfully revealed
substructures at different merging stages in other systems \citep[see,
for example,][]{dickey97,hba00,vangork03}.  We expect that spirals
already undergoing strong environmental effects present different
degrees of \hi\ deficiency, while the newcomers to the cluster
neighborhood display normal gas content.  In order to apply this
method to A\,85 we built a catalogue of the brightest blue member
galaxies\footnote{In the absence of morphological classification, we
expect that bright blue objects will very likely be late-type
galaxies.}  based on the following strategy. (1) Magnitude and color:
taking superCOSMOS data we produce a color-magnitude diagram to
exclude the red sequence.  In this fashion we select all galaxies
brighter than $b_{\rm J} = 18.0$ and bluer than $b_{\rm J} - r_{\rm F}
= 1.2$.  (2) Position: we take only the objects within the reported
region surveyed in \hi, as shown in Fig.\,\ref{blueglobal}.  (3)
Velocity: we only consider galaxies with known redshift, and within
the effective \hi\ velocity coverage 14,600--18,400\,\km; this
represents a velocity interval roughly centered on the systemic value
for A\,85 (16,600\,\km) and encompassing roughly three times the
cluster velocity dispersion.

There are 42 galaxies matching these criteria which constitute our
sample of the brightest blue members of A\,85; they are listed in
Table\,\ref{tab1}. After the running number (first Col.), the next
three Cols. list galaxy names and positions.  Galaxies with an
asterisk were detected in \hi\ and have reported gas masses of over
10$^9$\,\msolar. The symbol ``\#'' indicates galaxies with strong
\halpha\ emission \citep[as detected by][]{boue08}, and ``\#\#''
indicates a high probability of harboring an AGN, according to
coincident positions with X-ray point-like detections by
\citet{sivakoff08}.  A particular case is the blue \hi-rich galaxy
J0043017-094721 which, due to its projected position very close to
A85[DFL98]425, had remained previously uncatalogued as an independent
source.  These two objects are roughly coincident with the position
that
\citet{abell89} reported for the cluster A\,87, but are separated in
velocity by more than 10,000\,\km.  The position of J0043017-094721 was
determined in this work from a MEGACAM image taken at the 3.6-m CFHT
\citep{boue08}. In Col.\,5 we give the
optical radial velocity of the galaxies, obtained from the compilation
described in Sect.\,\ref{sec_data}.  The respective estimated error
(rms) in velocity is given in Col.\,6.  Col.\,7 gives the $b_{\rm J}$
SuperCOSMOS magnitude of the galaxies. Col.\,8 gives the $b_{\rm J} -
r_{\rm F}$ color.  Col.\,9 gives the projected distance to the cluster
center in Mpc.

One of the most important results of this paper is shown in
Fig.\,\ref{blueglobal}, where a very striking asymmetry in the
blue-galaxy distribution through the whole cluster volume is seen.
Crosses indicate the 42 brightest blue (cluster-member) galaxies
plotted over a composite view of A\,85 seen in optical and X-rays.  As
expected, the 12 \hi\ detections (encircled crosses) are within this
sample, and the remaining 30 constitute the category of gas deficient
(\hi-non detected) blue galaxies.  To check if this asymmetric
distribution is really a property of the blue members of A\,85, we
examined the distribution of non-emission line cluster galaxies taken
from a sample of 232 objects with SDSS spectra, which should be
representative of red members. As expected, these display a very
regular distribution, concentrated around the cluster center
\citep{caretta08}.

The spatial distribution of the \hi\ detections follows the same
asymmetric trend as blue galaxies: all but two are projected east of
the center of A\,85 where they are roughly distributed in two regions.
The first \emph{group} of five \hi\ detections is projected at an
average distance of $\sim$30\prim\ NE from the center of A\,85.  An
interesting feature to notice is that these five \hi-rich blue
galaxies (A85[DFL98]374/461/491/502 \& [SDG98]3114) display an
intriguingly low velocity dispersion of just 120\,\km, but are spread
(in projection) across a large area of $\sim$1.2\,Mpc.

The second group of \hi\ detections is projected SE of A\,85, around
the position of our detected \emph{SE} substructure (see
Fig.\,\ref{delta}).  Five galaxies are detected in \hi\ in this region
with at least four of them having higher probabilities of belonging to
a real subcluster when compared to those in the NE (see below). In
addition to these
\hi\ detections (A85[DFL98]323,486,347 \& J004301-094721), many other
(gas poor) blue objects appear in the SE quadrant. The \hi-rich
galaxies are projected, on average, some 35\prim\ SE of the center of
A\,85, along the extension of the X-ray filament (indicated by an
ellipse in the figure).  Similar to the five galaxies detected NE of
A\,85, these are also spread over a large area \citep[more than one
Mpc,][]{hba08}, and they form a grouping with low velocity dispersion
(see Table\,\ref{tab1}).

The asymmetry in the distribution of blue galaxies is better
quantified if we refer to the quadrants indicated in the figure.
SW~$\to$~NW~$\to$~NE~$\to$~SE gives the sequence from the poorest to
the richest in blue galaxies. Relative to the other quadrants, the SW
region is almost empty of blue galaxies, containing only four of these
objects (all \hi-poor).  In comparison, six blue objects are seen in
the NW and nine in the NE, where two and five \hi\ detections were
found respectively. By far the richest region in blue galaxies is the
SE, where 23 blue objects are distributed from the highest density
zone of the cluster ICM, going through the X-ray filament area
(indicated by an ellipse), and then spread around the position of the
\emph{SE} substructure.  Only five of those are detected in \hi. Most
of the blue galaxies projected on this region display velocities in
the range 15,000--16,300\,\km\ (Table\,\ref{tab1}), and several
display wide H$\alpha$ equivalent widths (larger than 25\AA).  The
presence of this indicator of star formation \citep[][and Paper II in
prep.]{boue08} strongly supports the hypothesis of a merging between a
subcluster of galaxies and A\,85.

%

\section{Discussion}\label{sec_discuss}

\subsection{Substructures across A\,85}

The results of the previous section clearly reveal the following substructures.

\noindent
{\bf The core:} The central region of the cluster appears to be
dynamically very active and presents strong signs of past activity
\citep{kempner02,durret05}.  We confirm this by detecting the most
prominent substructure (\emph{C2}) projected $\sim$5\prim\ west of the
cluster center and consisting of two blobs.  The ridge defined by
these two blobs is oriented in a SE-NW axis, and the position angle of
this axis is roughly the same as the orientation of the X-ray filament
and its extension to the \emph{SE} substructure.  Most of the galaxies
in \emph{C2} present velocities around 15,800\,\km, well under the
systemic value of A\,85 (16,600\,\km).

The southern part of \emph{C2} has been seen in X-rays as a diffuse
source with Chandra \citep[see Fig.\,1 of][the structure which they
call "the southwest subcluster"]{kempner02}. Here we report that the
clump of galaxies associated with this structure further extends some
10\prim\ to the N (see Fig.\,\ref{delta}), with another clump centered
on the second brightest galaxy of A\,85 (KAZ\,364).  This is a very
peculiar spiral reported as an individual X-ray source by
\citet{sivakoff08}, and classed an AGN. In spite of the large area
covered by this survey, the authors only report individual X-ray
sources; no diffuse emission is detected around KAZ\,364, equivalent
to that seen by
\citet{kempner02} in the southern tail of this grouping. This object also has the
lowest radial velocity in the whole A\,85 sample and, being outside
the VLA velocity coverage, it is not included in
Table\,\ref{tab1}. The $\delta$ signals associated with \emph{C2}
decreases if we remove this galaxy from the sample, but the
difference is not significant. The whole \emph{C2} structure, with
diffuse X-ray emission on its southern tail, strongly suggests a
subcluster undergoing an advanced merging process with the main cluster
core, in agreement with the general scenario proposed by
\citet{durret05}.

\noindent
{\bf The southern blob and filament:} Region \emph{SB}, projected some
10\prim\ south of the cD, probably consists of an almost disrupted
group of galaxies merging from the south
\citep{kempner02}. Intriguingly enough, the galaxies within this
structure display velocity values around 17,000\,\km, i.e. larger than
the substructures \emph{C2}, \emph{F} and \emph{SE}.  Clearly
identified in Fig.\,\ref{isodensity}, this substructure has the most
accentuated density enhancement, even comparable to the core of A\,85,
while its kinematics do not deviate much from the global pattern -
this is indicated with small green circles in Fig.\,\ref{delta} at the
same position.  The opposite occurs with clump \emph{F}, a
substructure projected midway along the X-ray filament described by
\citet{durret03}, about 20\prim\ away from the cluster center.  This
is not detected as a significant enhancement in the isodensity plot
(Fig.\,\ref{isodensity}) but deviates very much from the global
kinematics (see the large black circles in Fig.\,\ref{delta}).  This
structure is between the subgroup 3 in \citet{durret98b} and the
southern blob.

\noindent
{\bf The SE region:} As outlined in the previous sections, the SE is a
very interesting region. The substructure marked  \emph{SE} in
Fig.\,\ref{delta} is close to the projected position of the A\,87
cluster, roughly in the middle of the X-ray filament as detected in
the ROSAT PSPC field \citep{durret98b}, or equivalently, at the end of
the filament portion detected in the XMM-Newton field
\citep{durret03}.
This is a clump with very prominent kinematic deviations (large black
circles) but is barely detected as a galaxy density enhancement in
Fig.\,\ref{isodensity}.  The most likely explanation is that we are
dealing with a relatively poor subcluster undergoing a high velocity
infall towards A\,85. Many of the galaxies within this substructure
display velocities around 15,800\,\km, the same as substructures
\emph{F} and \emph{C2}. Most of the \emph{SE} galaxies have blue
colors, indicative of late type morphologies, and the galaxies
detected in \hi\ southeast of A\,85, being within this velocity range,
are certainly associated with this infalling group.  In Paper\,II we
analyse the spectral properties and NIR images of selected galaxies in
this area.

\noindent
{\bf The W group:} Another region with several large $\delta$ values
is seen about 35\prim\ west of the cluster center (marked with
\emph{W}). It is roughly at the same projected distance from the
cluster center as the \emph{SE} clump.  These objects are outside the
useful VLA-FOV and we have no information on their gas
content; furthermore it is outside all the reported X-ray surveys.  
This group of galaxies displays a velocity around
18,000\,\km. As this group is not projected along a major large scale
filament, it is impossible to say if a dynamical relation exists with
A\,85.

\noindent
{\bf The NE region:} The \hi-detected galaxies in the NE quadrant of
A\,85 (encircled crosses in Fig.\,\ref{isodensity}), in spite of
displaying a tight velocity dispersion, are neither detected in the
isodensity plot (Fig.\,\ref{isodensity}) nor as a kinematical
structure (Fig.\,\ref{delta}). We applied larger values of
$N_{neighb}$ unsuccessfully trying to separate these objects from the
global kinematics. This is likely due to the very large area over
which they are spread and to the small number of galaxies being part
of this possible structure.

%

\subsection{\hi-rich vs \hi-poor galaxies: the radial distribution}

Striking observational evidence found in many of the clusters
surveyed in \hi\ is the presence of galaxies and groups, projected at
large radius from the cluster center, that have been stripped of a
major fraction of their gas component 
\citep[e.g.][and references therein]{solanes01}.
Their large distance from the densest zones of the ICM raises the
question of whether the typical ram pressure stripping approach could
account for such effects. The best known examples have been reported
in Coma \citep{gavazzi87,hba00} and in Perseus\,I \citep{levy07}.
Based on Arecibo data, \citet{gavazzi87} reported a number of \hi\
deficient galaxies at large radii from Coma, projected on the
intermediate zone between this cluster and A\,1367. In addition to
this, \citet{hba00} reported one group of late-type galaxies projected
0.9\,Mpc east of the Coma center, with strong \hi\ deficiency. More
recently, \citet{levy07} found a very interesting picture in
Pegasus\,I, a system which lacks a harsh ICM and where we do
not expect ram pressure to strip large amounts of gas. These authors
report a number of spirals, near the cluster core but also in
foreground and background groups, ranging from moderate \hi\
deficiency  to truncated gas disks.  These cases can be added to other
stripped objects in low density environments, such as NGC\,4522 in
Virgo \citep[and references therein]{kenney04} and NGC\,2276 in the
NGC\,2300 group \citep{davis97}.  With the same trend we have A\,85,
with many gas-poor blue galaxies lying at 2-2.5\,Mpc from the cluster
center in projection.

A very important step forward on this issue was made by
\citet{tonnesen07}, who carried out a cosmological simulation of
cluster evolution. They found that gas stripping may occur out to the
cluster virial radius, as observed in A\,85 and the
clusters mentioned above. Very interestingly, their model predicts
that ram pressure stripping remains the most important mechanism
responsible for the loss of inter-stellar gas, and they found that a
fraction of galaxies is affected by this process in the intermediate
cluster zone (1.0--2.4\,Mpc) and up to the periphery (2.4--5.0\,Mpc).
This is in close agreement with our observational results.
With the help of Table\,\ref{tab1} and Fig.\,\ref{blueglobal} we see
that, among the 42 brightest blue member galaxies of A\,85, 30 are
deeply gas deficient and constitute our sample of blue gas-poor
galaxies. The radial distribution of these \hi-deficient objects is as
follows: only 8 are within the inner radius (r$<$\,1.0\,Mpc) while 22
are in the transition zone (1.0--2.4\,Mpc).  These gas-poor blue
galaxies along the transition zone of A\,85 would match the stripped
population predicted by \citet{tonnesen07}.

\begin{figure}
\resizebox{\hsize}{!}{\includegraphics[angle=-90,scale=.15]{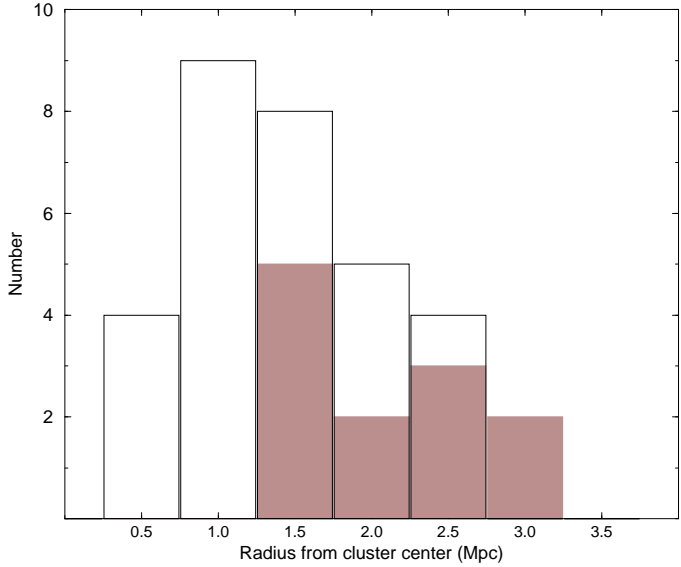}}
\caption{Histogram of the radial distribution of the two populations: 
\hi-rich (gray columns) vs \hi-poor (empty columns) galaxies. This clearly 
shows that, on average, the former are found at larger distances from the 
cluster center. }
\label{radial}
\end{figure}

Fig.\,\ref{radial} clearly shows that projected positions of \hi-rich
galaxies (filled columns), measured from the cluster center, follow a
different distribution when compared with the \hi-non detections
(empty columns, see also Table\,\ref{tab1}). The histogram displays
bins of 0.5\,Mpc, where the X-axis corresponds to the radius measured from
the cD galaxy. Average positions are 2.07\,Mpc and 1.42\,Mpc, for
\hi-rich and \hi-deficient galaxies, with standard deviations of 0.53
and 0.59 respectively. In order to examine the difference between both
distributions we applied the Kolmogorov-Smirnov test, which is
sensitive to global shape differences. We find they are different with
a significance level of 0.99. We complement this by applying the $t$
distribution to explore the difference of two means (case $n_1 + n_2 -
2$ = 40 degrees of freedom). The hypothesis $H_o: \mu_1 = \mu_2$ is
discarded with a confidence level of 0.99.

\subsection{Ram pressure stripping in A\,85}

In order to quantify the strength of the ICM--ISM interaction exerted
on our sample of blue galaxies around A\,85 we compute, as a first
approach, the cluster gas density distribution and extrapolate it to
radii of 2.0-2.5\,Mpc. We follow the hydrostatic-isothermal
$\beta$-model of \citet*{cavaliere76}, given by:

\begin{equation}\label{eq_rho}
\rho(r) = \rho_o [1+(r/r_c)^2]^{-3\beta/2}.
\end{equation} 

We take the parameters of the gas profile of A\,85 from \citet{chen07}: 
core radius $r_c$\,=\,82\,kpc, central density
$\rho_o$\,=\,0.0257\,cm$^{-3}$ (and assuming that this is an electrically neutral gas), 
and $\beta$\,=\,0.532. 
We then compute the ram pressure stripping
as a function of radius from the cluster center following 
\citeauthor*{gunn72}'s \citeyearpar{gunn72} equation: 
$P_{ram} = \rho_{ICM} \times v_{rel}^2$, where $\rho_{ICM}$ is the
density of the ICM at the galaxy position, and $v_{rel}$ is the galaxy
velocity relative to the ICM. Ideally, the latter should be the
component of the real velocity perpendicular to the galaxy disk
\citep[as defined e.g. in][]{cayatte94}, but not having information on
the inclination angle nor on the velocity of the galaxy in the plane
of the sky we will thus consider radial velocities, relative to the
cluster systemic movement, as a first approach. Normally, since galaxy
tangential velocity components are expected, this approximation
underestimates the real $v_{rel}$, making our value of $P_{ram}$ a
lower limit of the real one.


\begin{figure}
\resizebox{\hsize}{!}{\includegraphics[angle=-90,scale=.10]{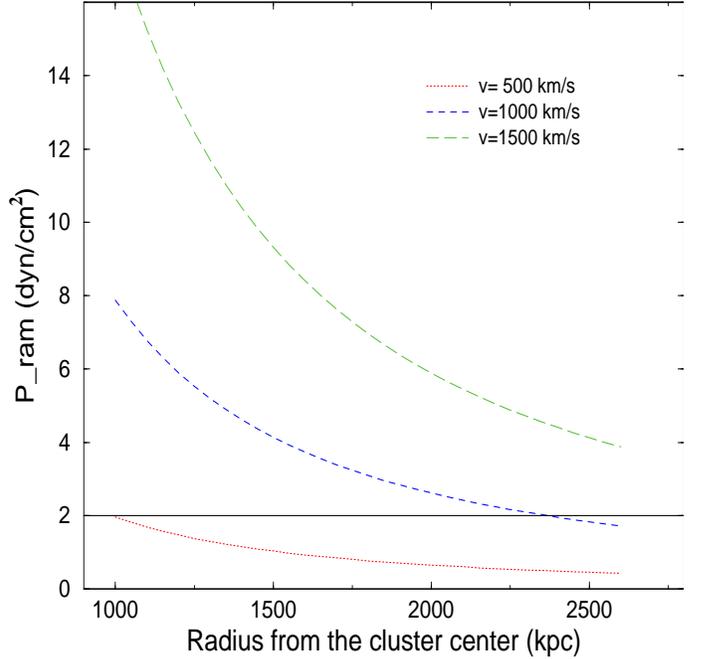}}
\caption{The ram pressure stripping, in units of 10$^{-12}$\,dyn\,cm$^{-2}$,
as a function of distance from the center of A\,85, estimated for
radii up to $\sim$2.5\,Mpc. Three different velocities, relative to
the ICM, were applied: 500, 1000 and 1500\,\km. As a comparison, the
solid line indicates the value of the restoring force F$_r$ (see
text).}
\label{P_ram}
\end{figure}

For the estimation of $P_{ram}(r)$ we consider three values of
$v_{rel}$: first we take 1000\,\km, roughly matching the (radial)
velocity dispersion of the cluster, which is also a lower limit of the
(radial) relative velocity between the \emph{SE} clump
($\sim$15,800\,\km) and A\,85 ($\sim$16,600\,\km). We include two
other velocity values as reference, 500\,\km\ and
1500\,\km. Fig.\,\ref{P_ram} displays the output of $P_{ram}$ for
values between 1.0 and 2.6\,Mpc.

In order to quantify the effects of ram pressure stripping on
individual galaxies we compare these values with the restoring force,
defined as $F_r\,=\,2\pi\,G\,\sigma_{tot}\,\sigma_{gas}.$ This is a
measure of the galaxy autogravitational binding force per cm$^{-2}$
(having units of dyn\,cm$^{-2}$, it can be directly compared with the
ram pressure). For this comparison we take the typical value of
$F_r\,=\,2\,\times\,10^{-12}\,dyn\,cm^{-2}$, obtained by
\citet{cayatte94} for spiral galaxies in Virgo. If we inspect
Fig.\,\ref{P_ram} we notice that, in spite of the low ICM density
values in the transition zone (where the \emph{SE} clump is found),
$P_{ram}$ is larger than the restoring force up
to 2.0\,Mpc from the cluster center for $v_{rel}\,\ge$\,1,000\,\km. 
Even the lowest velocity in our
estimation ($v_{rel}$\,=\,500\,\km) produces values of ram pressure of
the same order of magnitude as $F_r$ within the inner 1\,Mpc. 
This unexpected result confirms
that we should not discard ram pressure stripping as a mechanism
playing a major role in driving galaxy evolution when dealing with
high infall velocities relative to a cluster with a dense ICM such as
that of A\,85.

%

\section{Summary}\label{sec_sum}

Our results can be summarized as follows:

\begin{enumerate}
\item We first analysed the distribution of galaxies with measured
   redshift in the line of sight towards the complex A\,85/87/89,
   within a region of about 1.5 degrees side around the center of
   A\,85.  For A\,85 we find 367 member galaxies in the velocity range
   13,000--20,000\,\km.  We confirm the presence of two background
   structures near the position of A\,89. The farthest cluster in our
   sample is A\,87, at $\sim$39,000\km.

\item We revisited the 2-D
   (R.A., Dec.) distribution of galaxies through the whole A\,85/87/89
   complex and trace for the first time the isodensity contour map
   only with members of A\,85.  This highlights the substructures of
   A\,85 from the background clusters. Three major density peaks are
   revealed: two in the inner part of the cluster and one at
   $\sim$2\,Mpc in the SE direction, superimposed on the projected
   position of A\,87.

\item We ran the 3-D kinematical $\Delta$-test (R.A., Dec., $vel$) for
   the member galaxies of A\,85. We detect different clumps of
   galaxies going from the center of A\,85 and extending towards the
   SE. The first substructure reveals the inner structure of the core
   of A\,85, with a group of galaxies in an advanced merging stage
   with the main cluster body. We also report a structure coincident
   with the X-ray Southern Blob, roughly 0.7\,Mpc south of the center
   of A\,85.  Another substructure appears along the X-ray filament,
   more than 1\,Mpc SE of the cluster center. Finally, the \emph{SE}
   clump is projected at more than 2\,Mpc from the center of A\,85, at
   $\sim$15,800\km, coincident with the infalling sub-cluster revealed
   by the 2-D analysis.

\item We built a sample of the brightest blue galaxies, with 42
   members of A\,85 in the velocity range 14,600--18,400\,\km, and we
   trace their spatial distribution throughout the cluster. By taking
   into account the region surveyed in \hi\ and the velocity range
   covered by \citet{hba08}, we obtain information on the gas content of
   this sample, with just twelve objects (\hi-detections) showing
   rather normal gas content or mild deficiency; the remaining 30 blue
   objects represent deficient galaxies.  The spatial distribution of
   the whole 42 blue brightest members is totally unexpected, with
   about half of them projected on the SE quadrant of the
   cluster. This strongly supports the presence of sub-clusters in an
   early stage of merging with A\,85, infalling with a relative
   velocity greater than 1,000\,\km.

\item The distribution of the blue sample is also remarkable when measured as a function of
   the projected distance to A\,85. From the sample of 42 blue
   members, 32 are projected on the transition zone (1.0--2.4\,Mpc),
   and 22 of these are very gas deficient. We compare the positions of
   the two populations as a function of projected radius from the
   cluster center. We show that \hi-rich galaxies are projected, on
   average, much further than \hi-deficient ones: the mean projected
   distances are 2.1\,Mpc and 1.4\,Mpc respectively.

\item We estimated ram pressure stripping in A\,85 as a function
   of radius in order to quantify the environment's ability to
   decrease the gas fraction in spirals, in particular in the gas-poor
   blue galaxies reported in the SE region, $\sim 2$\,Mpc away from
   the center of A\,85.  Considering gas density parameters of A\,85
   ($\beta$ model and X-ray parameters) and different values of
   velocity relative to the ICM, we found that ram pressure is several
   times larger than the restoring force for typical spirals, when
   velocities relative to A\,85 are over 1000\,\km. This would explain
   the presence of the impressive sample of gas-poor blue-galaxies
   projected at the edge of the transition zone, i.e. at radius $\ga
   2$\,Mpc. The observational evidence shown here matches theoretical
   cluster models such as that of \citeauthor{tonnesen07}
   \citeyearpar{tonnesen07} who predict active gas stripping in that
   transition zone.

   \end{enumerate}

These results reveal an even more complex structure of A\,85 than
previously found, with groups of galaxies at different merging stages,
especially in the SE region but in the cluster core as
well. The observational evidence shown in this work confirms that the
cluster environment is playing an important role in driving galaxy
evolution. We show that ram pressure stripping might be effective up
to unusually large radial distances from a cluster center, when
infalling velocities are large enough. In a forthcoming paper (Paper
II) we will quantify the effects of gravitational mechanisms in
disturbing stellar disks, in particular in the SE region.

\begin{acknowledgements}
We thank the anonymous referee for her/his very useful comments which
helped us to strenghten and improve this paper. The authors are also
grateful to H. Andernach and E. Tago for providing radial velocity
data from their compilation.  HBA acknowledges CONACyT grant 50794,
and financial support from Institut d'Astrophysique de Paris and
Centro de Astrof\'{\i}sica da Universidade do Porto, allocated to
carry out working stays at these institutes.  CAC also aknowledges
support from CONACyT, grant 50921, and the University of Guanajuato
(DINPO) for the grant 0102/06. C. Lobo aknowledges finantial support
from project PTDC/CTE-AST/66147/2006 (FCT, Portugal).
\end{acknowledgements}
%

%
\bibliographystyle{aa} 
\bibliography{bravoalfaro_3035}
%
%

\begin{table*}
\caption{The sample of brightest blue member galaxies of A\,85. Objects with an asterisk are \hi\ detections. ``\#'' indicates strong \halpha\ emission, and
``\#\#'' indicates high probability of an AGN (see text).}  \label{tab1}
\centering
\begin{tabular}{rlcccrccrcc}
\hline\hline
Running & Name & 
$\alpha_{2000}$ & $\delta_{2000}$ & 
v$_{\rm opt}$ & $\Delta$v$_{\rm opt}$ &
$b_{\rm J}$ & $b_{\rm J}-r_{\rm F}$ &  Dist \\ 
number& & & & (\km) & (\km) & &  & (Mpc) \\ 
(1)              & (2)        & (3)        & (4)   & (5) & (6)    & (7) & (8)
\\ 
\hline 
1 & A85[DFL98]\,079*  & 0 40 31.66 & -09 13 20.0 & 15480 & 121 & 17.492 & 1.016 &
1.35 \\ 
2 & A85[DFL98]\,133   & 0 41 12.80 & -09 32 04.0 & 17156 & 114 & 16.547 & 1.028 &
1.12 \\ 
3 & A85[DFL98]\,139*  & 0 41 14.19 & -08 55 53.4 & 15146 & 185 & 17.978 & 0.593 &
1.60 \\ 
4 & A85[DFL98]\,150\# & 0 41 19.82 & -09 23 27.0 & 14728 & 125 & 17.649 & 1.104 &
0.62 \\ 
5 & A85[DFL98]\,167   & 0 41 27.16 & -09 13 43.1 & 14203 & 147 & 16.802 & 1.080 &
0.49 \\ 
6 & A85[DFL98]\,178   & 0 41 30.75 & -09 02 14.9 & 15231 & 116 & 17.210 & 1.092 &
1.11 \\ 
7 & A85[DFL98]\,193   & 0 41 34.91 & -09 00 47.5 & 17550 &  88 & 17.720 & 1.179 &
1.19 \\ 
8 & A85[DFL98]\,201   & 0 41 36.22 & -08 59 34.7 & 17969 & 174 & 17.432 & 1.121 &
1.26 \\ 
9 & A85[DFL98]\,221\#\#  & 0 41 42.96 & -09 26 21.4 & 16854 & 143 & 15.523 & 1.187 &
0.56 \\ 
10 & A85[DFL98]\,226\# & 0 41 45.44 & -09 40 33.3 & 17063 & 159 & 17.783 & 0.961 &
1.49 \\ 
11 & J0041490-095705   & 0 41 49.00 & -09 57 04.6 & 15536 &  90 & 16.860 & 0.968 &
2.59 \\ 
12 & A85[DFL98]\,255\# & 0 41 53.45 & -09 29 40.8 & 15326 & 115 & 15.757 & 0.528 &
0.77 \\ 
13 & A85[DFL98]\,267   & 0 41 57.34 & -09 35 23.9 & 15885 & 144 & 17.554 & 1.025 &
1.15 \\ 
14 & J0041599-093910   & 0 41 59.89 & -09 39 10.0 & 15761 & 113 & 16.330 & 0.833 &
1.41 \\ 
15 & A85[DFL98]\,279   & 0 42 02.16 & -09 40 17.6 & 14097 & 147 & 17.803 & 0.905 &
1.48 \\ 
16 & A85[DFL98]\,286   & 0 42 05.04 & -09 32 03.9 & 15832 & 168 & 15.959 & 0.969 &
0.96 \\ 
17 & A85[DFL98]\,291\# & 0 42 06.01 & -09 36 06.5 & 15865 & 117 & 17.476 & 1.198 &
1.22 \\ 
18 & A85[DFL98]\,300   & 0 42 10.28 & -08 55 51.6 & 17670 & 154 & 17.566 & 1.176 &
1.52 \\ 
19 & A85[DFL98]\,306   & 0 42 11.91 & -09 52 54.9 & 16296 & 125 & 17.160 & 1.114 &
2.34 \\ 
20 & A85[DFL98]\,316   & 0 42 17.27 & -09 28 47.7 & 15863 & 103 & 17.554 & 1.073 &
0.84 \\ 
21 & A85[DFL98]\,323*  & 0 42 18.72 & -09 54 13.5 & 15577 & 117 & 17.967 & 0.167 &
2.45 \\ 
22 & A85[DFL98]\,338   & 0 42 24.23 & -09 16 16.6 & 18184 & 115 & 17.237 & 1.080 &
0.58 \\ 
23 & A85[DFL98]\,347*  & 0 42 29.47 & -10 01 06.9 & 15191 &  85 & 17.750 & 0.595 &
2.93 \\ 
24 & A85[DFL98]\,356   & 0 42 33.76 & -08 52 53.5 & 17780 & 135 & 17.002 & 1.178 &
1.83 \\ 
25 & A85[DFL98]\,358   & 0 42 33.91 & -09 08 46.0 & 16685 & 157 & 17.351 & 1.048 &
0.96 \\ 
26 & A85[DFL98]\,363   & 0 42 34.75 & -09 39 18.1 & 16363 & 112 & 15.873 & 1.188 &
1.59 \\ 
27 & A85[DFL98]\,374*  & 0 42 41.47 & -08 56 49.0 & 15149 & 148 & 16.586 & 0.907 &
1.66 \\ 
28 & A85[DFL98]\,382\# & 0 42 43.91 & -09 44 20.7 & 15232 & 119 & 17.296 & 1.149 &
1.96 \\ 
29 & A85[DFL98]\,391   & 0 42 48.38 & -09 34 40.9 & 17934 & 175 & 17.885 & 0.945 &
1.46 \\ 
30 & J0043017-094721*  & 0 43 01.71 & -09 47 34.3 & 15110 &  22 & 17.411 & 0.269 &
2.29 \\ 
31 & A85[DFL98]\,429   & 0 43 04.16 & -09 32 43.1 & 15846 & 156 & 17.499 & 1.159 &
1.56 \\ 
32 & A85[DFL98]\,435   & 0 43 06.00 & -09 50 14.4 & 14767 & 135 & 16.953 & 1.196 &
2.48 \\ 
33 & A85[DFL98]\,447   & 0 43 10.91 & -09 40 53.9 & 16489 &  72 & 16.110 & 1.143 &
2.02 \\ 
34 & A85[DFL98]\,451\#\#  & 0 43 11.59 & -09 38 16.2 & 16352 & 103 & 16.024 & 1.054 &
1.90 \\ 
35 & A85[DFL98]\,461*  & 0 43 14.34 & -09 10 21.4 & 15008 & 142 & 19.005 & 0.444 &
1.49 \\ 
36 & A85[SDG98]\,3114* & 0 43 19.56 & -09 09 12.3 & 15060 &  22 & 19.360 & 0.285 &
1.60 \\ 
37 & A85[DFL98]\,483   & 0 43 30.48 & -09 43 59.4 & 15157 & 151 & 17.309 & 1.087 &
2.39 \\ 
38 & A85[DFL98]\,486*  & 0 43 31.16 & -09 51 46.4 & 16582 &  94 & 17.044 & 0.564 &
2.79 \\ 
39 & A85[DFL98]\,491*  & 0 43 34.04 & -08 50 37.0 & 14989 & 130 & 17.164 & 0.559 &
2.52 \\ 
40 & A85[DFL98]\,494   & 0 43 36.31 & -09 25 47.7 & 15051 & 195 & 16.830 & 1.168 &
1.83 \\ 
41 & A85[DFL98]\,496*  & 0 43 38.77 & -09 31 20.6 & 15068 & 130 & 17.165 & 0.504 &
2.00 \\ 
42 & A85[DFL98]\,502*  & 0 43 43.98 & -09 04 22.9 & 14830 & 128 & 17.694 & 0.772 &
2.10 \\ 
\hline 
\end{tabular}
\end{table*}

%

\end{document}